\documentclass[journal]{IEEEtran}
\usepackage{graphicx}
\usepackage{booktabs}
\usepackage{array}
\usepackage{cite}
\usepackage{url}
\usepackage[hidelinks]{hyperref}
\usepackage{balance}
\usepackage{subcaption}

\title{\huge Movable-Element STAR-RIS for 6G: From Programmable Propagation to Programmable Geometry}

\author{Wali Ullah Khan$^{*}$,~\IEEEmembership{Member,~IEEE}, Muhammad Adil$^{**}$ \\ $^*$Interdisciplinary Centre for Security, Reliability, and Trust (SnT), University of Luxembourg, Luxembourg \\ $^{**}$Department of Electronics Engineering, University of Rome Tor Vergata, 00133 Rome, Italy}

\begin{document}
\maketitle

\begin{abstract}
Reconfigurable intelligent surfaces (RISs) make the wireless propagation environment programmable, while simultaneously transmitting and reflecting RISs (STAR-RISs) extend this capability to users located on both sides of a surface. However, conventional STAR-RIS architectures retain a fixed physical geometry after deployment. Movable-element STAR-RIS (ME-STAR-RIS) introduces an additional spatial degree of freedom by allowing the surface elements to reposition within prescribed regions while maintaining electronic control of their transmission and reflection responses. This combination of electromagnetic and geometric reconfiguration can alter propagation distances, multipath combinations, spatial correlation, interference, near-field focusing, and sensing geometry. This article presents a system-level perspective on ME-STAR-RIS through the concept of programmable geometry. We discuss its operating principles, movement architectures, and promising applications in communications, security, near-field systems, sensing, and high-mobility networks. A representative case study comparing optimized fixed and movable STAR-RIS architectures illustrates measurable spectral-efficiency gains from limited local displacement and the resulting saturation behavior. Finally, key hardware, channel-acquisition, electromagnetic, energy, reliability, and control challenges are discussed toward practical ME-STAR-RIS deployment.
\end{abstract}



\begin{figure*}[!t]
\centering
\includegraphics[width=\textwidth]{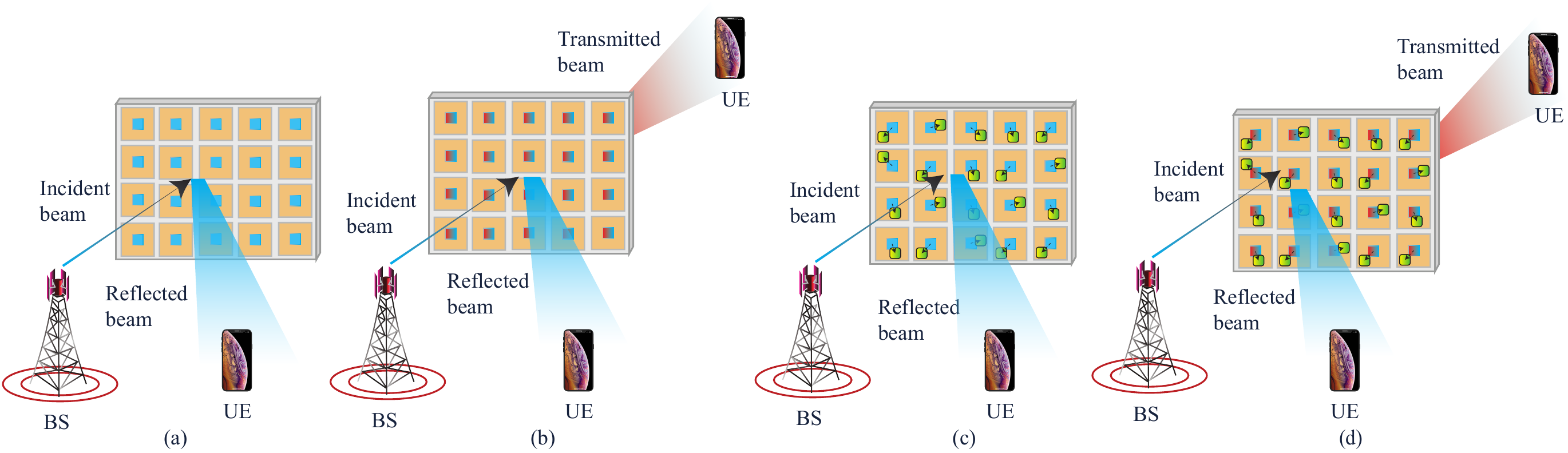}
\caption{Evolution toward programmable geometry: (a) conventional reflecting RIS, (b) fixed STAR-RIS, (c) movable-element RIS, and (d) movable-element STAR-RIS.}
\label{fig:evolution}
\end{figure*}

\section{Introduction}

Reconfigurable intelligent surfaces (RISs) have changed how wireless system designers view the propagation environment \cite{9424177,10584518}. Conventional wireless systems largely treat buildings, walls, scatterers, and obstacles as uncontrollable parts of the channel, while transmitters and receivers compensate through beamforming, coding, diversity, and scheduling. RIS technology introduces a different paradigm by embedding electronically controllable elements into the environment so that impinging electromagnetic waves can be redirected or reshaped \cite{11466360}. In this sense, part of the wireless propagation channel becomes a programmable network resource.

A conventional reflecting RIS, however, primarily manipulates signals in the half-space facing the incident wave. This limitation motivated simultaneously transmitting and reflecting reconfigurable intelligent surfaces, commonly referred to as STAR-RISs or STARSs \cite{10133841}. By controlling both reflected and transmitted components of an incident signal, STAR-RIS enables users located on both sides of a surface to be served and thereby extends programmable propagation toward full-space coverage. Existing STAR-RIS designs support operating strategies such as energy splitting, mode switching, and time switching, and have demonstrated significant flexibility for wireless coverage and multiuser transmission~\cite{10907789,11663258}.

Despite this increased electromagnetic flexibility, conventional RIS and STAR-RIS architectures retain an important constraint, i.e., their physical element locations remain fixed after deployment \cite{10138693}. Electronic tuning changes how an element responds to an incident wave, but it does not change the spatial point at which that wave is sampled. Consequently, the surface can program its electromagnetic response while its underlying geometry remains predetermined.

Movable-element technologies challenge this assumption by introducing controlled physical displacement as an additional spatial degree of freedom \cite{11386877}. Research on movable antennas and movable RISs has shown that relatively small positional changes can exploit spatial variations in the wireless channel~\cite{11611741,11205320}. Instead of adjusting only a complex coefficient at a predefined position, a movable element can select a more favorable location within a prescribed region. Such displacement can modify propagation distance, phase accumulation, multipath combinations, spatial correlation, and interference relationships. This creates the possibility of controlling not only how a surface interacts with an electromagnetic wave, but also where that interaction occurs.

Movable-element STAR-RIS (ME-STAR-RIS) combines these two forms of programmability. Each element retains its transmission and reflection capabilities while its physical position can additionally be adjusted within an allowable movement region \cite{10891163}. Recent studies have begun to establish the potential of this architecture. A general ME-STARS communication framework jointly considers active beamforming, STAR coefficients, and element positions under energy-splitting, mode-switching, and time-switching operation~\cite{zhao2026mestars}. Subsequent works have investigated ME-STARS for physical-layer security~\cite{zhao2026secure}, near-field wideband communications~\cite{zhu2025wideband}, and integrated sensing and covert communication~\cite{iscc2025movable}. These studies confirm the value of geometric reconfiguration, but the literature remains at an early stage and is still dominated by application-specific optimization problems.

The emerging literature therefore motivates a broader question, i.e., 
\emph{what changes when an intelligent surface can program not only its
electromagnetic response but also its geometry?}
This article addresses this question from a broader system-level perspective
through the concept of programmable geometry.
As illustrated in Fig.~\ref{fig:evolution}, RIS first made reflection
programmable, STAR-RIS extended this capability to simultaneous transmission
and reflection, and ME-STAR-RIS further makes the spatial configuration of
the surface controllable. We explain how electromagnetic and geometric reconfiguration complement each other, discuss practical movement architectures and operating timescales, and identify communication, security, near-field, sensing, localization, and high-mobility scenarios in which geometric adaptation can be particularly beneficial. A representative communication case study compares an electronically optimized fixed STAR-RIS with its movable counterpart to illustrate the spectral-efficiency benefit of limited local element displacement. Finally, we discuss the hardware, channel-acquisition, electromagnetic, energy, reliability, and control challenges that must be addressed before programmable geometry can evolve from a promising optimization concept into deployable wireless infrastructure.

\section{How Movable STAR-RIS Works?}
\subsection{Dual Electromagnetic and Geometric Reconfiguration}
The word ``movable'' can easily create the wrong impression. An ME-STAR-RIS does not necessarily mean that a large wall-mounted panel physically travels from one location to another. The more interesting architecture allows individual elements, small groups of elements, or reconfigurable subarrays to move within constrained regions while the overall surface remains installed at a fixed location. Movement may occur along tracks, within planar cells, on a flexible substrate, or through microelectromechanical positioning mechanisms. The important property is that the electromagnetic sampling locations of the surface are no longer permanently fixed.

This distinction leads to two complementary control domains. The first is the familiar electromagnetic domain. A STAR element can control the response of the incident wave through transmission and reflection coefficients, phase adjustment, and an appropriate operating protocol. Energy-splitting operation permits an element to contribute simultaneously to both transmission and reflection. Mode switching assigns an element to one side at a time, while time switching changes the operation over time. These principles remain relevant when the elements become movable; mobility does not replace STAR control.

The second domain is the geometric domain. Here, the controller decides where the elements should be positioned, how widely they may move, which movement directions are allowed, and how minimum spacing or mechanical constraints should be respected. The geometry may be reconfigured element by element or in groups. Movement may be continuous in a small planar region, restricted to horizontal or vertical tracks, or selected from a finite set of mechanically realizable positions. Recent near-field wideband ME-STARS research, for example, has studied region-based, horizontal, vertical, and diagonal movement patterns and found that different mobility structures lead to different performance--hardware tradeoffs~\cite{zhu2025wideband}.

\emph{Why is this useful when STAR-RIS already provides phase control?} The answer lies in the difference between modifying a signal at a fixed point and changing the channel itself. Electronic phase adjustment can compensate for propagation phase after the wave reaches a predefined element position. Physical displacement changes the propagation before the electronic coefficient is applied. An element that moves by a fraction of a wavelength experiences a different path length and therefore a different channel phase. In multipath environments, the relative phases of several arriving components also change, potentially transforming destructive combinations into constructive ones or vice versa. When distance-dependent attenuation varies across the movement region, the received amplitude can change as well.

This distinction becomes even more important when several links are involved. A STAR-RIS typically serves users on both sides of the surface, and the same element participates in channels toward multiple terminals. Moving one element modifies several cascaded links simultaneously. A location that improves one user may weaken interference toward another user or an eavesdropper. Conversely, a displacement that benefits the reflection region may alter the transmission-region channel in an undesirable way. ME-STAR-RIS design is therefore not simply ``phase optimization plus movement''; it is a coupled geometric control problem in which the best array layout depends on the objectives of the network.

Near-field propagation provides an especially compelling example. In a conventional far-field approximation, waves across an array are often represented mainly through direction-dependent phase progression. In the near field of electrically large surfaces, individual elements can experience noticeably different propagation distances and amplitudes. The geometry of the array then affects the location and shape of spatial focal regions. Repositioning elements can modify these distance-dependent relationships and may allow the surface to form more favorable focusing patterns. In wideband systems, where a configuration optimized for one frequency may perform poorly at another, geometric adjustment also provides another way to reduce frequency-dependent focusing errors. Recent ME-STARS work has specifically used element displacement to combat near-field wideband beam-squint effects and has reported advantages relative to fixed-position STARS~\cite{zhu2025wideband}.

\begin{figure}[!t]
\centering
\includegraphics[width=\columnwidth]{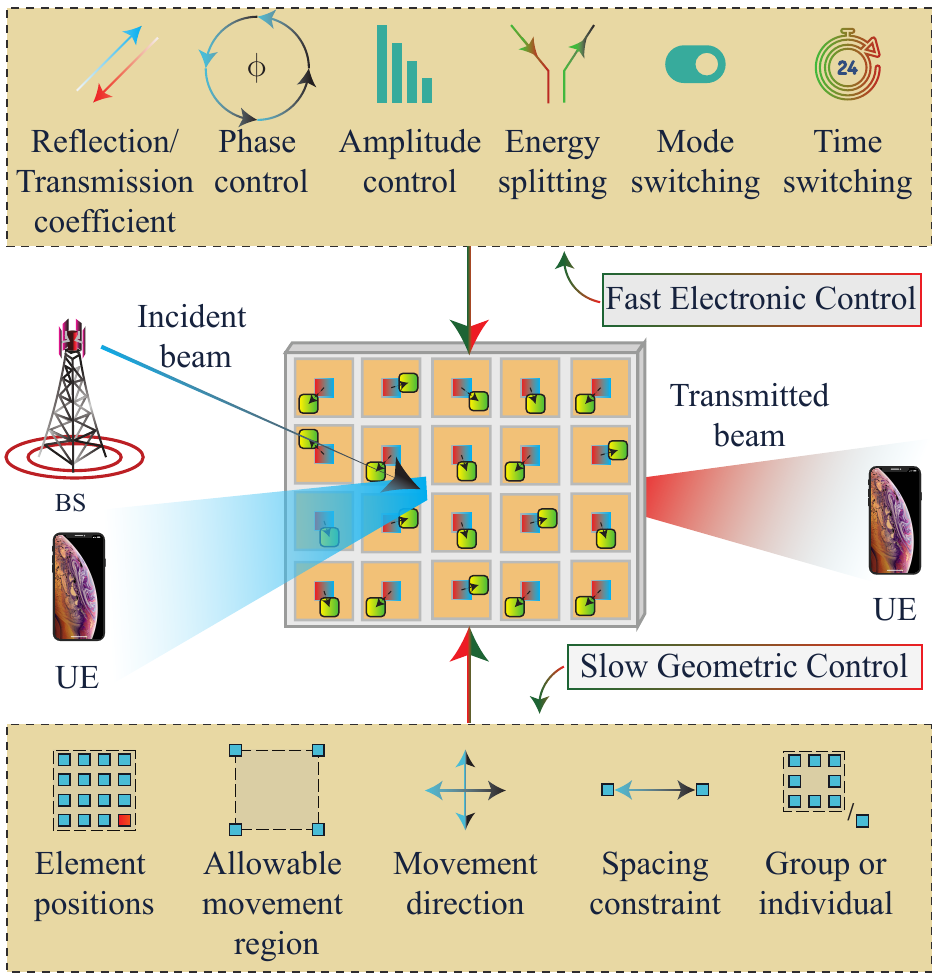}
\caption{Dual electromagnetic and geometric control in an ME-STAR-RIS.}
\label{fig:dualcontro}
\end{figure}

The resulting architecture can be viewed as a surface with dual programmability, as summarized in Fig.~\ref{fig:dualcontro}. Fast electronic variables control transmission, reflection, and phase. Slower geometric variables determine the spatial locations from which those responses are generated. This suggests that ME-STAR-RIS should not be understood as a replacement for existing STAR-RIS technology. Instead, it adds another layer on top of the existing electromagnetic control plane.

\subsection{Movement Architectures and Operating Timescales}
This dual-domain view also helps explain why movement should not necessarily occur at the same rate as electronic reconfiguration. Semiconductor-based phase control can react quickly to changing channel conditions, whereas mechanical motion is likely to be slower and more energy consuming. A practical surface may therefore update element positions only when the large-scale geometry, user distribution, blockage pattern, or propagation statistics change appreciably, while phase and transmission/reflection coefficients continue to track faster variations. Such two-timescale control may become the most practical operating principle for movable intelligent surfaces.

Another important design question concerns how much mobility is actually needed. Allowing every element to move freely over a large two-dimensional region maximizes geometric flexibility but introduces substantial mechanical complexity. Restricting each element to a narrow track, moving small groups rather than individual elements, or permitting only a few discrete positions can reduce actuator count and control overhead. The near-field ME-STARS literature already indicates that restricted movement modes can retain much of the mobility benefit while reducing hardware requirements. This observation suggests that the engineering objective should not be maximum movement freedom; it should be the minimum geometric flexibility required to capture most of the available channel gain.

\section{Where Can Programmable Geometry Matter Most?}
The potential value of ME-STAR-RIS depends strongly on the propagation environment and system objective. If the channel varies little across the allowable movement region, displacement may provide only marginal benefit. The concept becomes most attractive when small positional changes create sufficiently different communication, interference, or sensing conditions. Several emerging 6G scenarios naturally exhibit this property.

\subsection{Communications, Interference Management, and Security}
Physical-layer security provides a clear example. In a fixed STAR-RIS system, the controller adjusts active beamforming and surface coefficients to strengthen legitimate links while reducing information leakage. Yet the physical locations of the surface elements remain unchanged, so the geometric relationship among the transmitter, legitimate receivers, and passive eavesdroppers is fixed. Element relocation can modify these relationships. A position that preserves a strong legitimate channel may produce less favorable coherent combining toward an eavesdropper, providing additional spatial selectivity without adding RF chains or radiating power.

\begin{figure}[!t]
\centering
\includegraphics[width=\columnwidth]{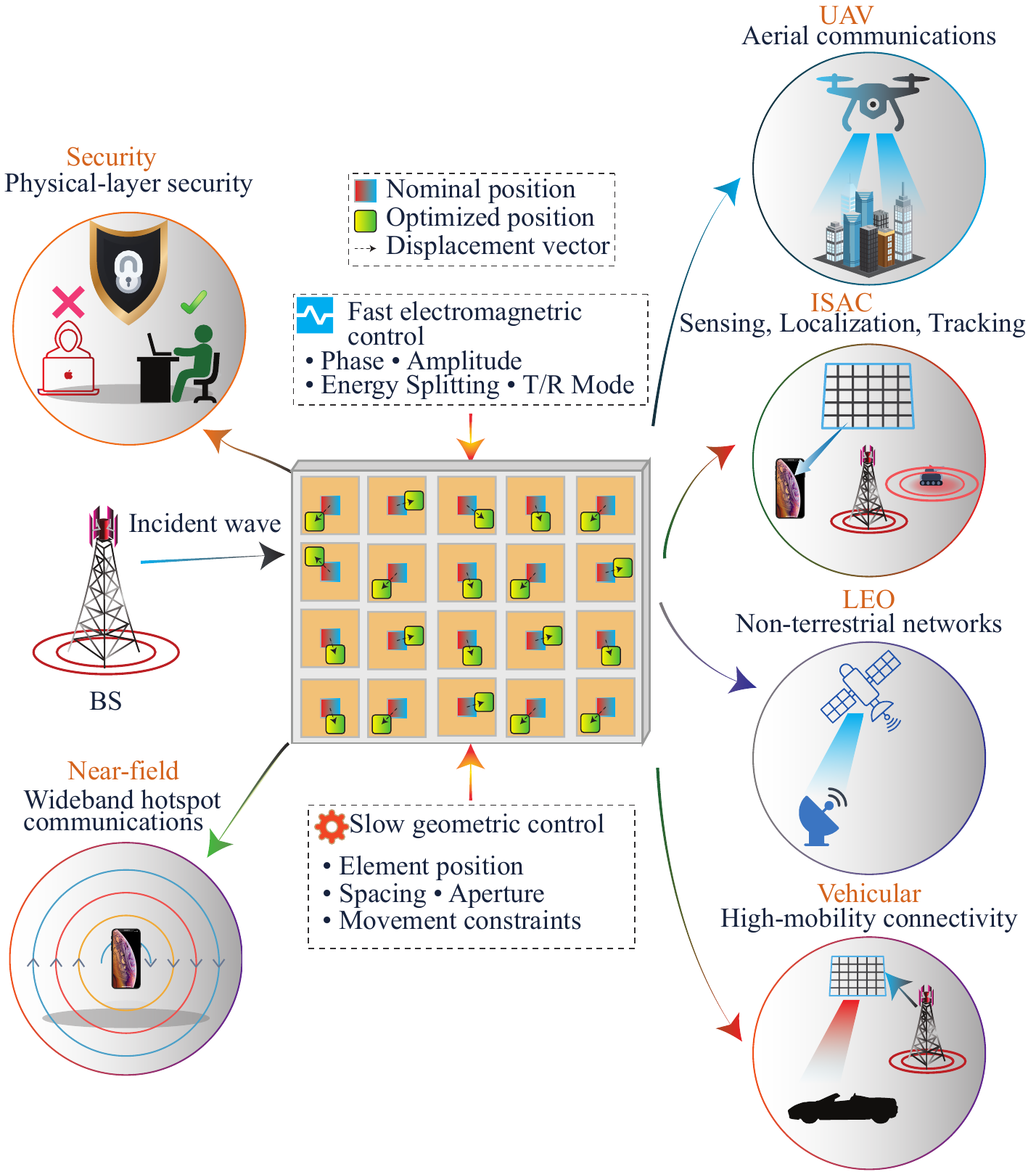}
\caption{Representative 6G application landscape and two-timescale operating concept for ME-STAR-RIS.}
\label{fig:applications}
\end{figure}

This idea has already been evaluated in recent ME-STARS security research. The reported results show approximately a 25\% secrecy improvement over conventional fixed-position STARS for the investigated scenario, while also showing that the benefit eventually saturates as the allowable movement region becomes sufficiently large~\cite{zhao2026secure}. The precise percentage should not be interpreted as universal; however, the qualitative behavior is highly informative, indicating that a relatively limited movement region may be sufficient to capture most of the available geometric diversity.

\subsection{Near-Field, ISAC, and Localization}
Integrated sensing and communication is another promising application because sensing and communication react differently to geometric changes. A STAR-RIS may simultaneously support users in different regions while redirecting energy toward sensing targets. Moving the elements changes not only the communication channels but also the illumination geometry, spatial focusing conditions, clutter relationships, and potentially localization sensitivity. A recent study has already investigated movable-element STAR-RIS for integrated sensing and covert communication under imperfect channel information~\cite{iscc2025movable}. The broader implication is that programmable geometry may help an ISAC system alter its physical aperture configuration according to the task. A communication-oriented configuration may favor spatial separation among user channels. A sensing-oriented configuration may instead favor target illumination or spatial resolution. A localization-oriented configuration could seek more informative geometric diversity. Whereas a fixed STAR-RIS changes the wavefront generated by one fixed aperture, an ME-STAR-RIS can change both the wavefront and the sampling geometry of the aperture itself.

\begin{table*}[t]
\caption{Emerging Research Directions in Movable-Element STAR-RIS}
\label{tab:related}
\centering
\renewcommand{\arraystretch}{1.15}
\begin{tabular}{|p{2.7cm}|p{3.1cm}|p{3.0cm}|p{3.0cm}|p{4.1cm}|}
\hline
\textbf{Representative work} & \textbf{Main scenario} & \textbf{Geometry control} & \textbf{STAR operation/design} & \textbf{Main message} \\
\hline
Zhao \emph{et al.}~\cite{zhao2026mestars} (2026) & Multiuser communication & Element-position optimization & ES, MS, and TS & Joint geometric and electromagnetic control improves weighted sum rate. \\ \hline
Zhao \emph{et al.}~\cite{zhao2026secure} (2026) & Physical-layer security & Confined element movement & Flexible STAR beamforming & Mobility improves secrecy and exhibits diminishing returns with movement range. \\ \hline
Zhu \emph{et al.}~\cite{zhu2025wideband} (2025) & Near-field wideband communication & Region, horizontal, vertical, and diagonal movement & Joint STARS/position design & Geometric adaptation can mitigate near-field beam squint. \\ \hline
 Zhou \emph{et al.}~\cite{iscc2025movable} (2025) & Sensing and covert communication & Movable STAR elements & Independent/coupled phase models & Mobility adds another communication--sensing design degree of freedom. \\
\hline
\end{tabular}
\end{table*}

Near-field wide-band communication provides another particularly compelling motivation. As carrier frequencies increase and apertures grow electrically larger, near-field operation becomes increasingly relevant. Conventional narrowband far-field beam steering becomes insufficient because different elements experience different propagation distances and because wideband beams can focus differently across frequencies. True-time-delay structures provide one approach but increase hardware complexity. The recent ME-STARS near-field study proposes element mobility as an alternative or complementary degree of freedom and demonstrates effective beam-squint mitigation. It also shows that different movement modes produce different balances between performance and implementation complexity~\cite{zhu2025wideband}.

\subsection{High-Mobility and Non-Terrestrial Networks}
Non-terrestrial, aerial, and vehicular systems provide another attractive setting. Satellite, UAV, and road-vehicle geometries change continuously, so a fixed intelligent surface must repeatedly compensate for changing propagation using only electronic coefficients. A movable surface introduces slower geometric adaptation on top of fast electronic tracking. For example, element positions could be adapted according to the long-term trajectory of a UAV, the dominant elevation direction of a satellite pass, the distribution of vehicles around a roadside surface, or persistent blockage patterns. Electronic STAR coefficients could then handle fast variations around that geometry.

This two-timescale viewpoint may be especially relevant to making ME-STAR-RIS practical. Attempting to mechanically reposition elements for every small-scale fading realization would likely be inefficient. Instead, movement could respond to more slowly evolving quantities such as user clusters, dominant paths, target directions, blockage states, or long-term channel statistics. Fast electronic reconfiguration would continue to handle instantaneous beamforming and interference control. In this architecture, mechanical mobility becomes a structural adaptation mechanism rather than a replacement for conventional real-time beamforming.

Fig.~\ref{fig:applications} groups these opportunities around a common two-timescale control view, in which fast electronic adaptation operates around a more slowly reconfigured geometry. ME-STAR-RIS may also interact naturally with localization, wireless power transfer, cell-free architectures, and edge intelligence. The common feature across these applications is that network performance depends not only on signal strength but also on geometric diversity and spatial relationships. A controllable aperture geometry can potentially create favorable measurement diversity for localization, concentrate energy differently for wireless power transfer, or adapt surface resources to changing distributed-user clusters.

Table~\ref{tab:related} summarizes the compact but rapidly expanding ME-STAR-RIS literature. The emerging literature remains concentrated on a small number of communication, security, near-field, and sensing scenarios, which leaves room for a broader system and deployment perspective. The general ME-STARS communication study established joint position and beamforming optimization under the principal STAR protocols~\cite{zhao2026mestars}. Secure ME-STARS extended the idea to full-space eavesdropping~\cite{zhao2026secure}. Near-field wideband ME-STARS investigated multiple movement modes and beam-squint mitigation~\cite{zhu2025wideband}. Movable-element STAR-RIS has also begun to appear in integrated sensing and covert communication~\cite{iscc2025movable}. Together, these works indicate a transition from demonstrating that movement can help to understanding where, how much, and at what implementation cost it should be used.

\section{Case Study: How Much Does Mobility Buy?}
A new degree of freedom is useful only if its gain survives a fair comparison against an electronically optimized fixed surface. To isolate the value of programmable geometry, we therefore consider a compact terrestrial case study in which the fixed and movable architectures use the same base-station antenna array, the same number of STAR elements, the same propagation realizations, and the same energy-splitting operation. The movable architecture receives only one additional capability, i.e., each element may shift locally around its nominal grid position.

\begin{figure*}[t]
    \centering
    \begin{subfigure}[t]{0.32\textwidth}
        \centering
        \includegraphics[width=\linewidth]{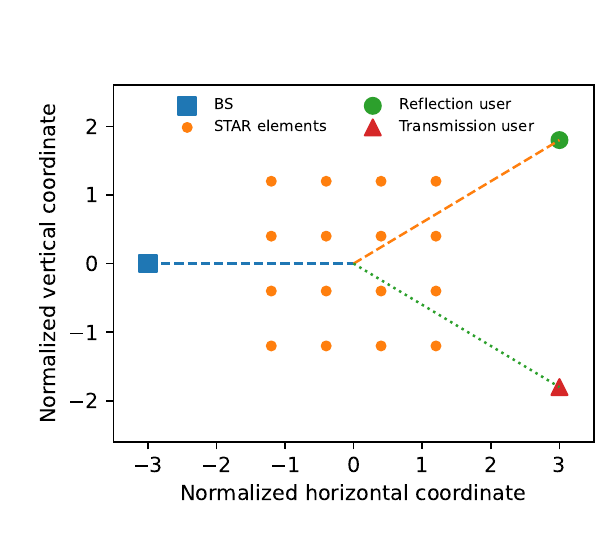}
        \caption{}
        \label{fig:case_geometry}
    \end{subfigure}\hfill
    \begin{subfigure}[t]{0.34\textwidth}
        \centering
        \includegraphics[width=\linewidth]{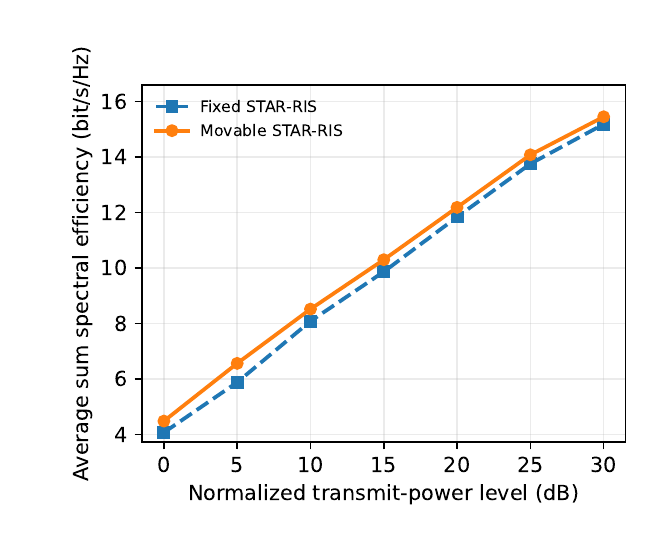}
        \caption{}
        \label{fig:case_power}
    \end{subfigure}\hfill
    \begin{subfigure}[t]{0.34\textwidth}
        \centering
        \includegraphics[width=\linewidth]{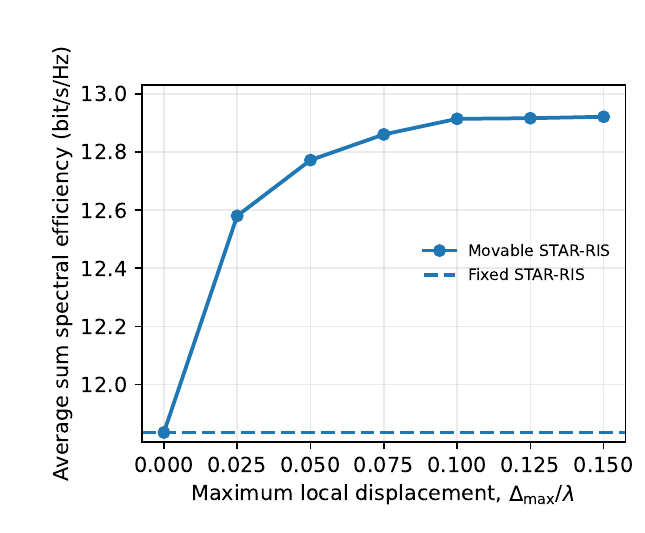}
        \caption{}
        \label{fig:case_movement}
    \end{subfigure}
    \caption{Illustrative ME-STAR-RIS communication case study:
    (a) two-user reflection/transmission geometry;
    (b) average sum spectral efficiency versus normalized transmit-power level;
    and (c) average sum spectral efficiency versus the maximum local element displacement.}
    \label{fig:case_study}
\end{figure*}

\subsection{Scenario and Fair Benchmarking}
The illustrative setup in Fig.~\ref{fig:case_study}(a) contains a four-antenna base station, a $4\times4$ STAR surface, one user in the reflection region, and one user in the transmission region. 

The surface begins from a regular grid with a nominal inter-element spacing
of $0.8\lambda$, where $\lambda$ denotes the carrier wavelength. Each movable
element is confined to a local square centered at its nominal position, while
a minimum pairwise inter-element separation of $0.5\lambda$ is enforced. The BS--surface and surface--user links contain a dominant component together with several weaker angular components so that different points inside a movement cell sample meaningfully different spatial channel combinations. A weak direct BS--user component is also retained.

For both architectures, the electronic design is optimized using the same
coordinate-search procedure. The BS uses regularized zero-forcing precoding,
while the STAR reflection and transmission phases and the energy-splitting
variables are updated to improve the two-user sum spectral efficiency. The
fixed benchmark performs these electronic updates with the regular element
grid held unchanged. The movable design starts from that optimized fixed
solution and then alternates the same electronic updates with feasible
position updates. Consequently, any performance difference is attributable
to the additional geometric degree of freedom rather than to a weaker
fixed-surface benchmark. The results are averaged over ten independently
generated channel realizations, with identical realizations used for the
fixed and movable configurations. Unless otherwise stated, the maximum local
displacement is set to $\Delta_{\max}=0.15\lambda$. The receiver-noise
variance is normalized to unity, and the transmit-power axis is reported in
a normalized dB scale, where 10~dB corresponds to one normalized transmit-power
unit.

\subsection{Spectral-Efficiency Gain Across Transmit Power}
Fig.~\ref{fig:case_study}(b) compares the average sum spectral efficiency
as the normalized transmit-power level is increased, with
$\Delta_{\max}=0.15\lambda$. Both architectures benefit from additional
power, but the movable surface remains above the fixed benchmark over the
complete range. The relative gain is most visible in the low-to-moderate-power regime, i.e., in this illustrative setup the average improvement is about 11.0\% at 0~dB and 12.6\% at 5~dB. The gain becomes smaller at higher power, falling to a few percent around 20--30~dB. This behavior is intuitive. When power is scarce, a better spatial configuration can materially improve how the available energy is converted into useful signal dimensions. At higher power, the electronically optimized fixed surface already supports high spectral efficiency and the incremental value of geometric refinement becomes more modest.

The important conclusion is therefore not that physical motion replaces transmit power or electronic beamforming. Instead, geometry acts as an additional resource that can improve the efficiency of the same aperture and power budget. The magnitude of that gain will vary with channel richness, direct-path strength, element count, and the allowed movement geometry, so the percentages in Fig.~\ref{fig:case_study} should be interpreted as a representative case study rather than universal ME-STAR-RIS gains.

\subsection{How Much Movement Is Enough?}
Fig.~\ref{fig:case_study}(c) addresses the practical question of movement range at a representative 20-dB power level. Starting from the optimized fixed grid, the local displacement bound is increased gradually. The average sum spectral efficiency rises quickly for very small displacements, i.e., allowing only $0.025\lambda$ local motion already produces an average gain of about 6.4\%. Increasing the displacement bound to $0.05\lambda$ raises the gain to about 8.0\%, while the improvement reaches roughly 9.1\% by $0.10\lambda$. Beyond that point the curve is nearly saturated in the considered geometry.

This result has a direct hardware implication. A useful ME-STAR-RIS may not need elements that travel over large distances. Small local movement cells can capture much of the available spatial diversity, after which additional mechanical range delivers little benefit. The relevant engineering target may therefore be the smallest displacement range that captures most of the geometry gain, rather than the largest physically possible movement range. Such a design reduces actuator travel, repositioning time, mechanical wear, and calibration burden while preserving most of the communication benefit.

Taken together, the three panels illustrate two broader design lessons. First, programmable geometry is complementary to conventional electromagnetic control rather than a substitute for it. Second, the value of mobility is strongly nonlinear in the movement range, i.e., the first small amount of displacement can be far more valuable than extending an already adequate movement region. These observations motivate hardware architectures based on limited local motion and slow geometry adaptation rather than unrestricted continuous movement.

\section{From Theory to Deployable Surfaces}

The conceptual principle of movable-element STAR-RIS is simple: jointly control the electromagnetic response and the surface geometry. Realizing this principle in a practical system, however, is considerably more challenging. The value of ME-STAR-RIS will ultimately depend not only on the achievable communication gain, but also on whether the required movement can be implemented with acceptable latency, energy consumption, calibration burden, and hardware reliability.

\subsection{Hardware, Energy, and Reliability}

The first challenge is physical actuation. Most existing ME-STAR-RIS studies assume that each element can move continuously within a prescribed region, whereas practical hardware will impose restrictions on displacement range, positioning resolution, movement direction, and actuation speed. Possible realizations may rely on microelectromechanical mechanisms, miniature tracks, deformable substrates, mechanically reconfigurable subarrays, or other forms of programmable metasurface geometry. These constraints suggest that future designs should increasingly consider hardware-aware mobility models rather than unrestricted element movement.

Mechanical reconfiguration also changes the energy-efficiency argument traditionally associated with passive RIS technology. Although a movable element may consume little energy while remaining stationary, repeated repositioning of a large number of elements can introduce non-negligible actuation cost. Consequently, geometry should not be updated whenever a marginally better channel is available. A more practical strategy is to trigger movement only when the expected performance benefit justifies the associated actuation, calibration, and control overhead.

Reliability is equally important. Frequent movement may introduce mechanical wear, positioning errors, and long-term calibration drift. This motivates restricted movement architectures in which element groups move together, predefined tracks are used, or a finite codebook of tested geometric configurations is maintained. Such approaches reduce mechanical complexity while preserving part of the geometric diversity offered by fully movable elements.

\begin{table*}[t]
\caption{From Theoretical Mobility to Practical ME-STAR-RIS Deployment}
\label{tab:roadmap}
\centering
\renewcommand{\arraystretch}{1.18}
\begin{tabular}{|p{3cm}|p{6.4cm}|p{7.4cm}|}
\hline
\textbf{Challenge} & \textbf{System implication} & \textbf{Promising direction} \\ \hline
Mechanical latency
& Geometry cannot follow fast fading
& Slow geometry adaptation with fast electronic STAR control \\
\hline
Actuation energy
& Mobility changes the energy-efficiency argument of passive surfaces
& Event-triggered repositioning and movement-aware energy optimization \\
\hline
Channel acquisition
& Channels vary continuously with position
& Geometry-based models, sensing-assisted maps, and digital twins \\
\hline
Position uncertainty
& Small errors can cause significant phase mismatch
& Closed-loop position sensing and online calibration \\
\hline
Mutual coupling
& Irregular spacing changes electromagnetic response
& Coupling-aware channel and optimization models \\
\hline
Mechanical reliability
& Repeated movement introduces wear
& Group movement, restricted tracks, and geometry codebooks \\
\hline
Optimization complexity
& Position, beamforming, and STAR coefficients are coupled
& Model-driven learning and low-dimensional geometry control \\
\hline
Network integration
& Mechanical states introduce new control overhead
& Hierarchical and two-timescale network control \\
\hline
\end{tabular}
\end{table*}

\subsection{Channel Acquisition and Two-Timescale Control}

Channel acquisition becomes more demanding when the surface geometry itself is variable. A conventional STAR-RIS already requires knowledge of the transmitter--surface and surface--user channels. With movable elements, these channels also vary with position, making exhaustive estimation over all candidate locations impractical.

Geometry-aware channel models can reduce this burden by exploiting path angles, propagation distances, dominant scatterers, and user locations to predict how the channel evolves as the elements move. Sensing-assisted channel maps and digital twins may further support this process by combining environmental knowledge with online measurements. Instead of estimating every possible configuration directly, the controller can infer promising geometric states from a lower-dimensional representation of the propagation environment.

The different response times of electronic and mechanical reconfiguration naturally lead to a two-timescale architecture. Fast STAR coefficients can adapt to instantaneous channel variations, while the physical geometry is updated more slowly according to user distribution, dominant propagation paths, blockage states, or mobility trends. In this way, element movement acts as a structural adaptation mechanism rather than attempting to follow every small-scale fading realization.

This operating principle is particularly attractive in dynamic scenarios. A roadside surface, for example, could adapt its geometry to slowly varying traffic patterns while electronically tracking individual vehicles. Similarly, in UAV or NTN systems, the surface geometry could be configured according to a predicted trajectory or elevation range, with fast electronic control compensating for shorter-term fluctuations.

\subsection{Electromagnetic-Aware and AI-Assisted Control}

Another important challenge is electromagnetic realism. Moving the elements changes not only their propagation distances and phases but potentially also their mutual coupling, impedance, radiation patterns, and local electromagnetic interactions. These effects become increasingly important when the surface forms irregular or tightly spaced configurations. Communication-theoretic geometry optimization should therefore evolve toward coupling-aware and hardware-informed models, potentially supported by reduced-complexity surrogates derived from full-wave electromagnetic simulations.

The resulting control problem is highly coupled because beamforming, STAR coefficients, and element locations must be coordinated under both electromagnetic and mechanical constraints. AI can help reduce online complexity, but its most useful role is likely to be model-assisted rather than purely data driven. Learning methods can predict when movement is worthwhile, identify low-dimensional geometric configurations, estimate spatial channel maps, or provide fast approximations to computationally expensive optimization routines.

Digital twins, model-driven learning, and graph-based architectures are particularly promising because they can incorporate known physical constraints such as movement limits, minimum element spacing, and transmission/reflection operation while learning the difficult parts of the environment dynamics. The long-term objective should therefore not be unrestricted element mobility, but a practical form of programmable geometry that captures most of the available performance gain with limited movement, low control overhead, and reliable hardware operation.

Table~\ref{tab:roadmap} summarizes the main deployment challenges and corresponding research directions. Overall, the key question is no longer whether movable elements can outperform fixed ones under idealized optimization. The more important issue is whether most of the useful geometric gain can be obtained using mechanically simple, energy-aware, and slowly reconfigured architectures suitable for real wireless networks.

\section{Conclusion}

Movable-element STAR-RIS extends programmable wireless environments by adding geometry as an additional controllable resource. Unlike conventional STAR-RIS, which electronically adjusts transmission and reflection from fixed locations, ME-STAR-RIS can also reposition its elements to reshape the propagation environment. The emerging literature and the case study presented in this article indicate that this extra spatial degree of freedom can improve communication, security, sensing, and near-field operation, while even relatively small local movements may capture a meaningful part of the available gain. The key challenge is now to translate this flexibility into practical hardware. Promising implementations will likely rely on limited and energy-aware element movement, slow geometry adaptation combined with fast electronic STAR control, and geometry-aware channel acquisition and optimization. The evolution of intelligent surfaces can therefore be viewed as a progression from programmable reflection, to full-space programmable transmission and reflection, and ultimately to programmable geometry.

\balance
\bibliographystyle{IEEEtran}
\bibliography{references}
\end{document}